# EXPERIMENTAL IMPROVEMENT OF 750-KEV H- BEAM TRANSPORT AT LANSCE ACCELERATOR FACILITY

Yuri K. Batygin*, LANL, Los Alamos, NM 87545, USA\

*Abstract*

The 750-keV low-energy beam transport of the Los Alamos Neutron Science Center (LANSCE) linac consists of two independent beam lines that facilitate simultaneous injection of $H^+$ and $H^-$ beams into the linear accelerator. The efficiency of beam transport within this structure is largely controlled by the effects of space-charge neutralization. In this paper, we report a series of experiments that were performed to determine the level, as well as the required timescale for the onset of beam space charge neutralization. Measurements performed along the structure indicate significant variation of neutralized space charge beam dynamics in the beamline, and the corresponding results have been used to substantially improve beam transport with diminished beam losses. The analysis of experimental and theoretical data employed in optimization of beam performance in the transport line is discussed.

## INTRODUCTION

Since commencing operation in 1972, the LANSCE Accelerator Facility has undergone a series of transformations and upgrades, reflecting the continually changing demands of its user-base. Currently, LANSCE utilizes four 800 MeV H- beams, as well as one 100 MeV proton beam. The most powerful 80 kW H- beam is accumulated in the Proton Storage Ring and is extracted to the Lujan Neutron Scattering Center facility for production of moderated neutrons with MeV-keV energy. Another H- beam, as a sequence of short pulses, is delivered to the Weapons Neutron Research Facility to create un-moderated neutrons in the keV- MeV energy range. The third H- beam is delivered to the Proton Radiography Facility and the fourth one to Ultra-Cold Neutron facility. Finally, the 23 kW proton beam is used for isotope production in the fields of medicine, nuclear physics, national security, environmental science and industry.

In 2014, the operation pulse rate of the LANSCE accelerator facility changed to 120 Hz, following a decade-long period of operation at a 60 Hz pulse rate. While working at 60 Hz with 825 μs pulse length, the H- source was operated with a 5% duty factor (60 x 0.000825). The source steadily delivered a 16 mA beam within a 28-day operation cycle, reaching a 0.53 A-hour lifetime (0.016 A x 28 days x 24 hours x 0.05 = 0.53 A-hour). Currently, the H- source operates at the highest-ever possible duty factor of 10%, delivering maximum current of 14 mA within 28 days with a 0.94 A-hour lifetime. After transformation to the double pulse rate, the beam power for the Weapons Neutron Research Facility increased by a factor of 2.5, while the power of other beams remains unaltered. Consequently, increased ion source lifetime with reduced peak current has demanded re-evaluation of tuning procedures in order to meet new challenges in beam operation.

---

* batygin@lanl.gov

# H⁻ ION SOURCE AND 750 KEV BEAM TRANSPORT

The LANSCE H⁻ beam injector includes a cesiated, multicusp-field, surface - production ion source [1], and a two-stage low-energy beam transport line (see Fig. 1). In the first stage, the extracted beam is accelerated up to 80 keV, and is then transported through a solenoid, electrostatic deflector, a 4.5° bending magnet, followed by a second solenoid. Subsequently, the 670 kV Cockroft-Walton column accelerates the beam up to an energy of 750 keV. The 750 keV LEBT consists of a quadrupole lattice, 81° and 9° bending magnets, slow-wave chopper, RF bunchers, an electrostatic deflector, as well as diagnostics and steering magnets, to prepare the beam for injection into the Drift Tube Linac (DTL). Slit-collector beam emittance measurements at 750 keV are performed at five locations: (1) TBEM1 (just after the Cockroft -Walton column), (2) TBEM2 (downstream of the chopper), (3) TBEM3 (downstream of the 81° bend before the RF pre-buncher), (4) TBEM4 (between the first RF (pre)-buncher and second (main) buncher), and (5) TDEM1 (before the entrance to the DTL).

Figure 2 illustrates the LANL surface-production H⁻ ion source [2]. Negative ions are created as a result of charge exchange at a molybdenum surface converter, in the presence of a thin layer of cesium. The generated H⁻ particles are then accelerated towards the extraction aperture. Correspondingly, beam emittance of this type of source is estimated as the phase space area comprised by a converter with radius $R_{conv}$, extractor aperture with radius $R_a$, and distance $L$ between them (admittance of source) [3]

$$э = \frac{4}{\pi} \frac{R_{conv} R_a}{L} \quad . \tag{1}$$

Normalized beam emittance $\varepsilon = \beta\gamma \, э$ is therefore given by

$$\varepsilon = \frac{4}{\pi} \sqrt{\frac{2eU_{conv}}{mc^2}} \frac{R_{conv} R_a}{L} \, , \tag{2}$$

where $U_{conv}$ is the voltage between the converter and the source body. In the LANSCE H⁻ ion source, $R_{conv}$ =1.9 cm, $R_{ext}$ = 0.5 cm, $L_{conv}$ =12.62 cm, $U_{conv}$ = 300 V, which yields a normalized beam emittance of $\varepsilon$ = 0.076 π cm mrad. This quantity is close to the experimentally observed value of four-rms normalized beam emittance $4\varepsilon_{rms}$ = 0.0703 π cm mrad (see Fig. 3). Accordingly, the ratio of total emitance to rms emittance for this source is $\varepsilon_{total}/\varepsilon_{rms} \approx 7$, which indicates that the beam distribution is between "Water Bag" and parabolic distribution in 4D phase space [4].

From Eq. (2) it follows that the reduction of converter voltage results in a reduction of normalized beam emittance [5]. Correspondingly, experimental lowering of extraction voltage from a nominal value of 300 V to 177 V resulted in a noticeable reduction of normalized beam emittance (see Table 1). Lowering of converter voltage is limited by the voltage (~ 170V) between heated filament cathode and source body for plasma generation. We note however, that maintenance of the extraction voltage at a nominal value of 300 V leads to a reduced frequency of source arc-downs, and thus provides long-term stability of source operation.

After acceleration in the 750 keV Cockroft -Walton column, the beam is transported through a 12 m transport channel, which contains 22 quadrupole lenses. Acceptance of the channel is given by [6]

$$A = \frac{\mu_o}{S(1+\upsilon_{max})^2} a^2 \, , \tag{3}$$

where $\mu_o$ is the phase advance of transverse oscillations at distance $S$, $a$ is the aperture of the channel, and $\upsilon_{max}$ is the envelope ripple factor determined by

$$R_{max} = \overline{R}(1+\upsilon_{max}), \tag{4}$$

where $R_{max}$ is the maximum value of envelope, and $\overline{R}$ is the average value of envelope. Figure 4 illustrates the dynamics of a single particle, together with beam envelopes corresponding to the maximum possible emittance (acceptance of the channel). As seen, a single particle performs one complete oscillation within a distance of 6.2 m. Maximum value of beam envelope is equal to beam pipe radius $R_{max} = 2.54$ cm, and average envelope is $\overline{R} \approx 1.5$ cm, therefore the ripple factor is $\upsilon_{max} \approx 0.7$. The average value of aperture along the channel is $a \approx 2$ cm. Based on these numbers, the acceptance of the channel is estimated according to Eq. (3) to be $A \approx 13.8 \ \pi$ cm mrad. Numerical simulations give a more accurate value of $A \approx 10 \ \pi$ cm mrad (see Fig. 4). Normalized acceptance of the channel is therefore

$$\varepsilon_{ch} = \beta\gamma A = 0.4 \ \pi \text{ cm mrad}. \tag{5}$$

Acceptance of the channel is 23 times larger than the rms beam emittance. Space-charge limited beam current in focusing channel is determined by [6]

$$I_{max} = \frac{I_c}{2}\frac{\mu_o}{S} A(\beta\gamma)^3 [1-(\frac{\varepsilon}{A})^2], \tag{6}$$

where $I_c = 4\pi\varepsilon_o mc^3/q = 3.13 \cdot 10^7$ [Amp] is the characteristic beam current. Based on the obtained numbers, the space-charge limited current in beamline is

$$I_{max} = 100 \ [1-(\frac{\varepsilon}{A})^2] \ [mA]. \tag{7}$$

A typical value of beam current extracted from LANL ion source within the last 10 years of operation is 13 – 17 mA. Average effective beam current due to space charge neutralization is around 30% of the actual beam current (see below). Equation (6) for limited beam current can be rewritten as

$$I_{max} = \frac{j_o \varepsilon_{ch}}{2}[1-(\frac{\varepsilon}{\varepsilon_{ch}})^2], \tag{8}$$

where $j_o$ is the normalized value of phase space density of the channel:

$$j_o = (\beta\gamma)^2 \frac{I_c \mu_o}{S} = 0.5 \ \frac{A}{cm \ mrad}. \tag{9}$$

Adopting an operation beam current of $I = 5$ mA, and a normalized rms emittance of $\beta\gamma \ \varepsilon_{rms} = 0.02 \ \pi$ cm mrad, the phase space density of the beam is

$$j_b = \frac{I}{4\beta\gamma \ \varepsilon_{rms}} = 0.0625 \ \frac{A}{cm \ mrad}. \tag{10}$$

The ratio $b_o = j_b / j_o$ is the space charge parameter, which determines the significance of space charge on beam dynamics [6]. In particular, a beam with $j_b > j_o$ is strongly affected by space charge, while the effect of space charge forces is negligible for a beam with $j_b \ll j_o$. For this beamline, the space charge parameter is $b_o = 0.125$, therefore the beam transport is emittance dominated. Correspondingly, the space charge depression factor is

$$\frac{\mu}{\mu_o} = \sqrt{1+b_o^2} - b_o = 0.88 . \tag{11}$$

However, space charge forces must be taken into account for precise beam matching and minimization of beam losses along the beamline. From Eqs. (9) - (11), the transported beam current is expressed through beam and channel parameters as

$$I = \frac{I_c}{2} \frac{\mu_o}{S} \ni (\beta\gamma)^3 [\frac{1-(\mu/\mu_o)^2}{(\mu/\mu_o)}] . \tag{12}$$

For maximum achievable beam current, space charge depression factor is equal to ratio of maximum emittance of the beam transported without losses, to acceptance of the channel, $\mu/\mu_o = \ni /A$ [6], and Eq. (12) is transformed into expression for maximum beam current, Eq. (6).

## SPACE CHARGE NEUTRALIZATION OF H⁻ BEAM

Ionization of residual gas by transported particles is an important factor of low-energy beam transport. Ionization of residual gas results in creation of electron-ion pairs, which neutralize the space charge of the primary beam. As an outcome of neutralization, an effective beam current, $I_{eff}$, is smaller than actual beam current, $I_{beam}$, which affects beam dynamics. Another important parameter is an effective beam emittance, $\varepsilon_{eff}$, which describes phase space area occupied by the beam after completion of neutralization process. Values of effective current and effective emittance are required for adequate description of the beam by envelope model, which is used for selection of quadrupole gradients and beam matching in beamline. Determination of $I_{eff}$, $\varepsilon_{eff}$ is achieved through study of time-dependent space charge neutralization process.

The value of space charge neutralization, $\eta$, is defined by

$$\eta = 1 - \frac{I_{eff}}{I_{beam}} . \tag{13}$$

Figure 5 illustrates the variation of beam emittance along beam pulse measured at TBEM4 and TDEM1 emittance stations. Figs. 6 illustrate the dependencies of Twiss parameters ($\alpha$, $\beta$) four - rms normalized emittance $4\varepsilon_{rms}$, and rms beam sizes $\sigma$ versus beam pulse length $\tau$ at TBEM4. Note that values of beam parameters are observed to stabilize after 200 µs.

Figure 7 illustrates a typical spectrum of residual gas, measured by a Residual Gas Analyzer (RGA) installed in the middle of the transport channel. The main components are $H_2$ (48%), $H_2O$ (38%) and $N_2$ (10%). The fraction of other components is significantly lower. Pressure measured by four ion gauges (IG) distributed along the channel is $P_{IG} = 5 \cdot 10^{-7}$ Torr. It is well known that RGA can be used mostly for determination of relative concentration of gases, while ion gauges give a more accurate value of total pressure [7]. Additionally, all ion gauges are calibrated for $N_2$ as the only gas in the system. Contribution of other gases to the total pressure is estimated by sensitivity coefficients $r_i$ (see Table 2). Total actual pressure in the system, $p_{act}$, based on combination of fractional contribution, $\chi_i$, and ion gauge reading, $P_{IG}$, is determined as [7]

$$p_{act} = \frac{P_{IG}}{\sum \chi_i r_i} \quad . \tag{14}$$

Individual partial pressures are then the product of fractional contribution of each gas and actual gas pressure $p_i = \chi_i p_{act}$. The number of molecules of each gas per unit volume, $n_i$, at pressure $p_i$ and absolute temperature $T$, is determined from the ideal gas law:

$$n_i = \frac{p_i}{k_B T}, \tag{15}$$

where $k_B = 1.38 \times 10^{-23}$ J·K$^{-1}$ is the Boltzman constant. Time required for ionization of residual gas by the incoming beam with velocity of $v = \beta c$ is determined by [8]:

$$\tau_n = \frac{1}{v \sum_i (\sigma_i n_i)}, \tag{16}$$

where $\sigma_i$ is the ionization cross section of the beam on gas component $i$ [9], and $v$ is the beam velocity. Combining Eqs. (14) - (16), the neutralization time is

$$\tau_n = \frac{k_B T}{v P_{IG}} \frac{\sum_i (r_i \chi_i)}{\sum_i (\sigma_i \chi_i)} \quad . \tag{17}$$

Taking into account the ion guage readings, the estimated neutralization time according to Eq. (17) is $\tau_n \approx 170.5 \ \mu s$. These time constants are close to the experimentally observed values as illustrated by Fig. 6.

A series of beam emittance scans along the 750 keV H$^-$ beam transport line were performed to determine the time and level of space charge neutralization of the beam, value of effective beam current under space–charge neutralization, and the value of effective beam emittance. Measurements were performed with an ion source pulse length of 825 μs, and were done between each pair of emittance stations TBEM1–TBEM2, TBEM2–TBEM3, TBEM3–TBEM4, TBEM4–TDEM1. The emittance was sampled within the last 50 μs of the ion source pulse. The beam pulse start time was varied between $\tau$ = 10 – 575 μs before the emittance sampling through delay in the 80 kV electrostatic deflector. A typical value of H$^-$ beam current at 750 keV was 14 – 17 mA.

Determination of the value of compensated space charge by residual gas ionization was done through comparison of measurement results and simulations, using the macroparticle code BEAMPATH [10] and envelope code TRACE [11]. In order to avoid ambiguity in determination of effective current and emittance, the study was done as a sequence of two steps. At the first stage of the simulations, the beam distributions measured at the starting station were reproduced in a BEAMPATH macroparticle model, and adopted as the initial distribution for subsequent beam simulations (see Fig. 8). At this step, no assumptions about beam emittance are made, because macro-particle distribution reproduces actual measured beam distribution. After that, simulations were performed between two emittance stations with variable beam current. At the subsequent measurement station we compared equivalent beam ellipses obtained from measurement and from simulation (see Fig. 9), and calculated the mismatch factor between them $F = 0.5(F_x + F_y)$, where

$$F_x = \sqrt{\frac{1}{2}(R_x + \sqrt{R_x^2 - 4})} - 1, \qquad (18)$$

and $R_x = \beta_{exp}\gamma_s + \beta_s\gamma_{exp} - 2\alpha_{exp}\alpha_s$ is the parameter indicating overlap of $x$ - beam ellipses. In the above expression, Twiss parameters obtained from experiment are denoted $\alpha_{exp}$, $\beta_{exp}$, $\gamma_{exp}$ and those derived from simulations are designated $\alpha_s$, $\beta_s$, $\gamma_s$. The same procedure is repeated for $F_y$. The smallest value of the mismatch factor $F$ determines the value of effective beam current under space-charge neutralization, $I_{eff}(F_{min})$.

At the second stage of analysis, the same procedure was repeated with the envelope code TRACE using different beam emittances, with the value of effective beam current obtained from the macroparticle model (see Fig. 10). A minimum value of the mismatch parameter indicates an effective value of beam emittance representing the beam in the envelope model in the most adequate way together with determined value of effective current. Let us note, that reversing of order of determination of effective current-emittance, or using envelope code only leads to uncertainty in results, because minimization of mismatching factor in this case might be obtained with various combinations ($I_{eff}$, $\varepsilon_{eff}$).

Figures 11 - 12 illustrate the results from the space-charge neutralization study between TBEM4 – TDEM1, utilizing the described method. Specifically, Figure 11 shows the value of mismatch factor $F$, Eq. (18), as a function of beam current in BEAMPATH simulations. Note that at the beginning of beam pulse, the mismatch factor is minimized at the largest value of beam current. This indicates the absence of space-charge neutralization. With longer beam pulses, the minimum of mismatch factor moves towards a smaller effective current, which corresponds to 50% – 60% space charge neutralization. Obtained values of effective beam current for each pulse length were used in TRACE code with different values of beam emittance (see Fig. 12). Minimum mismatch indicates the most appropriate combination of effective beam current and effective beam emittance in the envelope model.

The described method was used for space charge neutralization study of the H⁻ beam along the whole transport channel. Figure 13 illustrates the value of space charge neutralization, $\eta$, versus beam pulse length for the rest of the beamline. Space charge neutralization reaches a value of 50% between TBEM1-TBEM2, 100% between TBEM2-TBEM3, 80% - 90% between TBEM3-TBEM4, and 50% - 60% between TBEM4-TDEM1. Analysis of beam emittance (see Fig. 14) indicates that effective beam emittance in beam transport is close to the value of $\varepsilon_{eff} = 3.5\varepsilon_{rms}$. In the regions TBEM2-TBEM3, TBEM3-TBEM4 the value of beam emittance could be determined only when space charge neutralization is significantly below 100%, which corresponds to a beam pulse length of $\tau$ < 150 µs. Otherwise, when the effective current is close to zero, transformation of the beam ellipse from one point to another is independent on the value of beam emittance.

## BEAM EMITTANCE GROWTH

Two main sources were identified for beam emittance growth: RF bunching and beam chopping. Transverse beam emittance growth crossing a single RF gap is [12]

$$\frac{\Delta\varepsilon}{\varepsilon} = \frac{k^2\beta_w^2}{10}(\Delta\Phi)^2 \sin^2\varphi_s, \qquad (19)$$

where $\beta_w$ is the value of beam beta Twiss parameter in RF gap, $\Delta\Phi$ is the phase length of the bunch, $\varphi_s$ is the synchronous phase of the bunch, and $k$ is the parameter of RF gap with applied voltage $U$ and transit-time factor $T$:

$$k = \frac{\pi q U T}{mc^2 (\beta\gamma)^3 \lambda} . \tag{20}$$

In both the prebuncher and the main buncher, the beam radius is $R_w = 0.4$ cm, so the value of beta - Twiss parameter for a beam with unnormalized emittance $э_{rms} = \varepsilon_{rms}/(\beta\gamma) = 0.5\ \pi$ cm mrad is

$$\beta_w = \frac{R_w^2}{4\ э_{rms}} \approx 0.8\,m . \tag{21}$$

In the prebuncher, the RF voltage is $U = 4$ kV, while the main buncher voltage is $U = 13$ kV, with a transit time factor in both bunchers equal to $T = 0.68$. The bunch length in prebunchers is $\Delta\Phi = 2\pi$, which gives emittance growth in the prebuncher of $\Delta\varepsilon/\varepsilon = 2.2\%$. In the main buncher, $\Delta\Phi \approx 4.5$ rad, and the estimated emittance growth is $\Delta\varepsilon/\varepsilon = 15\%$. The experimentally observed RF induced beam emittance growth is within 20% (see Fig. 15 a, b).

The H⁻ chopper is located downstream from the Cockcroft-Walton column. It consists of two traveling-wave helix electrodes, which apply a vertical kick to the beam. The chopper is normally energized so that no beam gets through. An electrical pulse of length $\tau = 36$ ns and 290 ns for short and long beam pulses respectively, travels along the chopper allowing the unchopped part of the beam pulse to pass through. The minimum width of the chopper pulse is determined by the chopper risetime, which is about 10 ns.

The long bunch pulses feed the chopper at the rate of 2.8 MHz, which corresponds to one pulse every 358 ns, which is the revolution time for the Proton Storage Ring at the LANSCE Accelerator Facility. Following the chopper, the leading and trailing edges of the beam are bent vertically, which affects the beam emittance. Emittance growth due to the short chopper pulse can be as high as 60%, while for the longer pulse it is within a few percent (see Fig. 15 c, d). This is explained by the fact that emittance distortion is caused by the edges of chopper pulse with a rising-falling time of 10 ns. These edges comprise a significant fraction of the 36 ns pulse length, but are much smaller than the 290 ns pulse.

Emittance growth due to space charge is estimated as [13]

$$\frac{\varepsilon_f}{\varepsilon_i} = \sqrt{1 + (\frac{\mu_o^2}{\mu^2} - 1)(\frac{W_i - W_f}{W_o})} , \tag{22}$$

where $(W_i - W_f)/W_o$ is the "free-energy" parameter due to space-charge induced beam intensity redistribution. Parameter $(W_i - W_f)/W_o$ has a value of 0.01126 for "Water Bag" distribution, and 0.0236 for parabolic beam distribution in 4D phase space. The expected space charge emittance growth is therefore between 1.6% - 3.4%, which is close to the noise level in emittance measurements. In experiments, no noticeable emittance growth due to space charge was observed. Space charge plays an insignificant role in beam emittance growth because beam transport is emittance-dominated, but is an important factor in beam matching.

## BEAM MATCHING

In this work, we have reported an experimental survey of beam space charge neutralization in the LANSE accelerator facility. The performed study creates a basis for precise beam tuning within the structure. Beam matching is based on determination of beam emittances in both vertical and horizontal planes, and a judicious selection of the values of quadrupole lenses, in order to provide the required beam ellipse parameters at certain points through solution of envelope equations. The values of focusing fields are varied to minimize the mismatch factor $F$ (given by Eq. 18) between expected and calculated beam ellipses. Along the beam transport, there are several points where the beam has a waist: at the middle of chopper, at the entrance to prebuncher, as well as at the entrance of Main Buncher (see Fig. 16). Tuning is based on emittance scans at certain emittance stations; prediction of downstream quadrupole magnets setup which provides required Twiss parameters at certain points of beam transport; emittance scan at the next emittance station; recalculation of beam transport back to previous emittance station and correction of quadrupole setup. Matching from point to point requires adjustment of four Twiss parameters $\alpha_x, \beta_x, \alpha_y, \beta_y$, which is achieved by employing a combination of quadrupoles between them. Final matching at the entrance of Drift Tube Linac requires a beam ellipse with Twiss parameters $\alpha_x$ = 0.0271, $\beta_x$ = 26.01 cm, $\alpha_y$ = -0.0575, $\beta_y$ = 5.95 cm, determined by lattice structure of DTL.

Historically, it was assumed that the H$^-$ beam is close to being completely space-charge neutralized in the beam transport, and the effective current was assumed to be 1 mA. With the completed study of space-charge neutralization described above, it was found that the value of effective current differs significantly from the expected value of 1 mA (see Table 3). Table 4 illustrates beam transmission through the beamline tuned using the historically adopted method with an effective beam current of 1 mA, and that with the new approach based on our space-charge neutralization study. Accordingly, it was demonstrated that beam transmission through the beamline has improved significantly with the latter approach. Particularly, the optimization reported herein gives rise to the possibility of attaining the same peak output current of 10 mA, with a concurrent decrease of peak source current from 16 mA to 14 mA. The outcome of this improvement is that the Lujan Neutron Scattering Center maintains operation with nominal beam power of 80 kW, while the WNR beam has increased its current by a factor of 2.5.

## SUMMARY

A study of 750 keV H$^-$ beam transport was performed. Basic results can be summarized as follows. Twiss parameters and the beam size are functions of pulse length (stabilized after 200 μs). Beam emittance is also dependent on pulse length, indicating 10-20 % variation in the beginning of beam pulse. Furthermore, our space charge neutralization study indicates that the beam has various levels of neutralization (from 50% to 95% depending on position along beamline). Typical neutralization time is close to the analytical estimations of 170 μs. Application of envelope model shows that the ratio of effective beam emittance to rms emittance is close to $\varepsilon_{eff} / \varepsilon_{rms} \approx 3.5$. The performed analysis has granted the possibility to provide the 80 kW H$^-$ beam to LANSCE Lujan Center facility with significant increase of pulse rate of other LANSCE beams.

## ACKNOLEDGEMENT

Author is indebted to Lawrence Rybarcyk, Chandra Pillai and Gary Rouleau for help in performing of experiments and useful discussion of results.

Table 1. Beam emittance extracted from ion source as a function of converter voltage.

| Converter Voltage (V) | Analytical estimation of beam emittance ($\pi$ cm mrad) | Experimental $4\varepsilon_{rms}$ beam emittance ($\pi$ cm mrad) |
|---|---|---|
| 300 | 0.076 | 0.072 |
| 177 | 0.058 | 0.060 |

Table 2. Ionization of residual gas.

| Gas | Ionization cross-section $\sigma_i$ ($10^{-20}$ m$^2$) | Fractional Contribution to Total Pressure, $\chi_i$ | Ion Gauge Gas Sensitivity $r_i$ | Product $\chi_i r_i$ | Product $\sigma_i \chi_i$ ($10^{-20}$ m$^2$) |
|---|---|---|---|---|---|
| H$_2$ | 0.504 | 0.48 | 0.46 | 0.22 | 0.242 |
| H$_2$O | 1.78 | 0.32 | 1.12 | 0.36 | 0.569 |
| N$_2$ | 1.9 | 0.097 | 1.0 | 0.097 | 0.184 |
| O$_2$ | 1.99 | 0.011 | 1.01 | 0.011 | 0.0218 |
| Ar | 1.85 | 0.032 | 1.29 | 0.041 | 0.0592 |
| CO$_2$ | 2.75 | 0.048 | 1.4 | 0.067 | 0.132 |
| Total | | 1.0 | | 0.796 | 1.208 |

Table 3. Effective beam current along beam transport.

| Region | $I_{eff} / I_{beam}$ |
|---|---|
| TBEM1 – TBEM2 | 0.56 |
| TBEM2 – TBEM3 | 0.08 |
| TBEM3 – TBEM4 | 0.16 |
| TBEM4 – TDEM1 | 0.50 |

Table 4. Transmission of chopped H$^-$ beam before and after improvement. Reduction of beam intensity (80%) between TBCM1 and TBCM2 is due to beam chopper.

| Current Monitor | Beam current (mA) before improvement | Beam current (mA) after improvement |
|---|---|---|
| TBCM01 | 15.72 | 13.88 |
| TBCM02 | 11.00 | 11.1 |
| TBCM03 | 10.2 | 10.6 |
| TBCM04 | 9.88 | 10.52 |
| TBCM05 | 9.80 | 10.44 |
| TDCM01 | 9.72 | 10.42 |

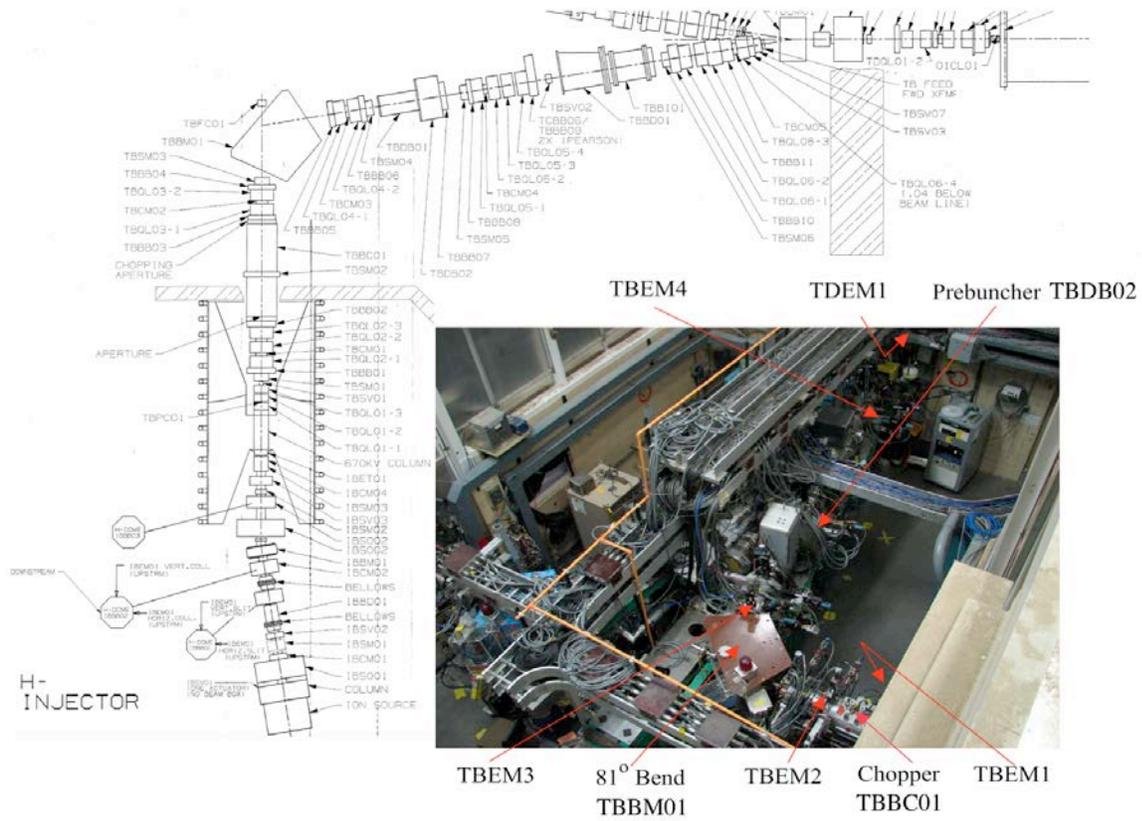

Figure 1: Layout of LANSCE H- injector.

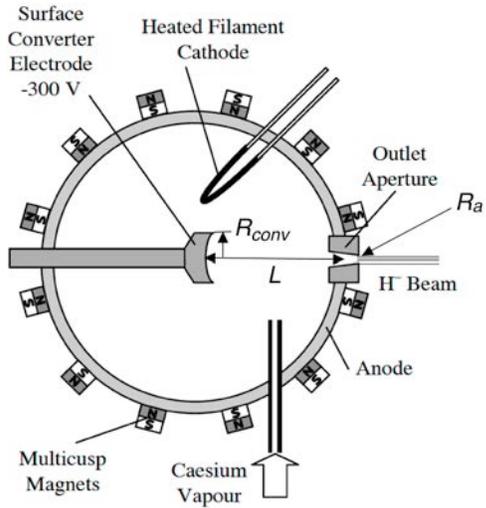

Figure 2. LANL H- Ion Source [2].

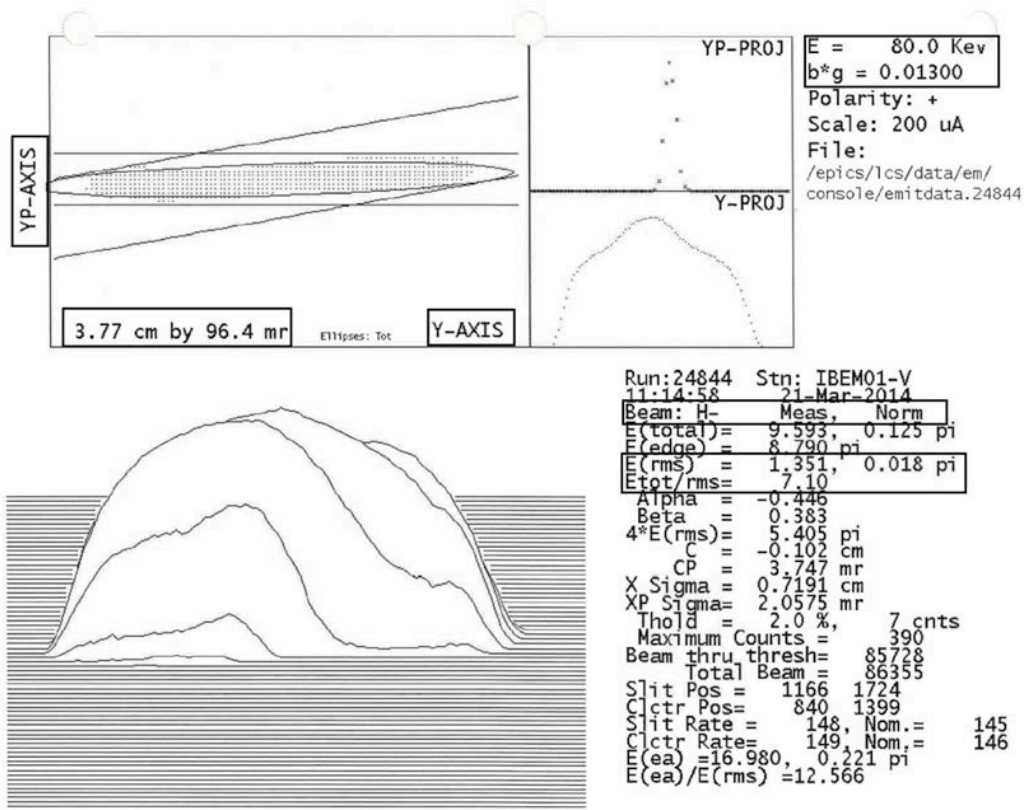

Figure 3. Emittance of H- beam extracted from LANSCE ion source.

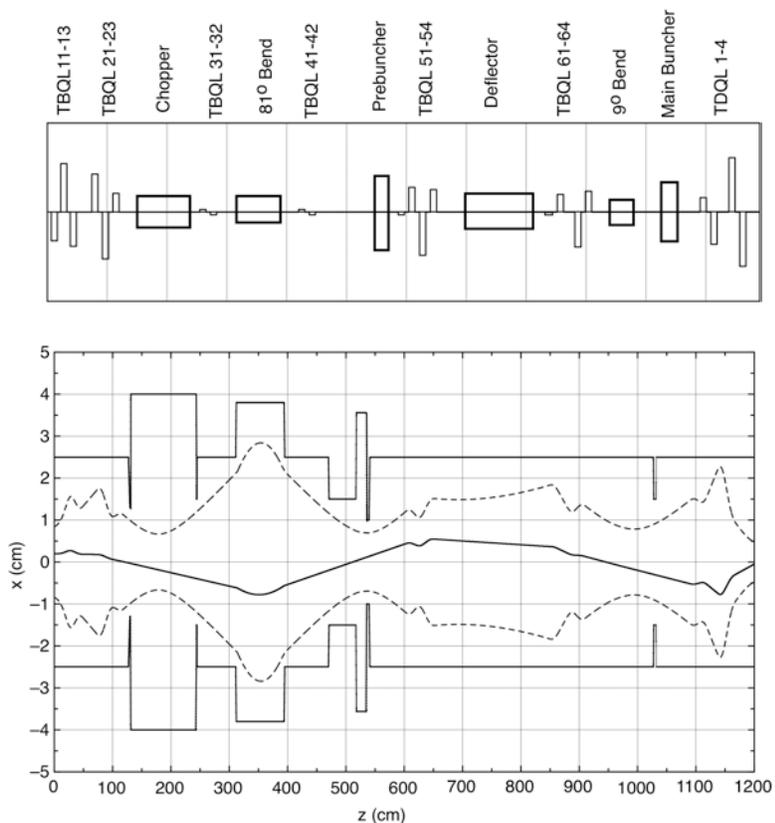

Figure 4. Beam dynamics in 750-keV H- beam transport: (solid) single trajectory, (dotted) envelope of the beam with maximum possible emittance (acceptance of the channel) A = 10 $\pi$ cm mrad.

τ = 10 μs

τ = 100 μs

τ = 300 μs

τ = 500 μs

Figure 5. Measured horizontal beam emittance at (left) TBEM4 and (right) TDEM1 along beam pulse length.

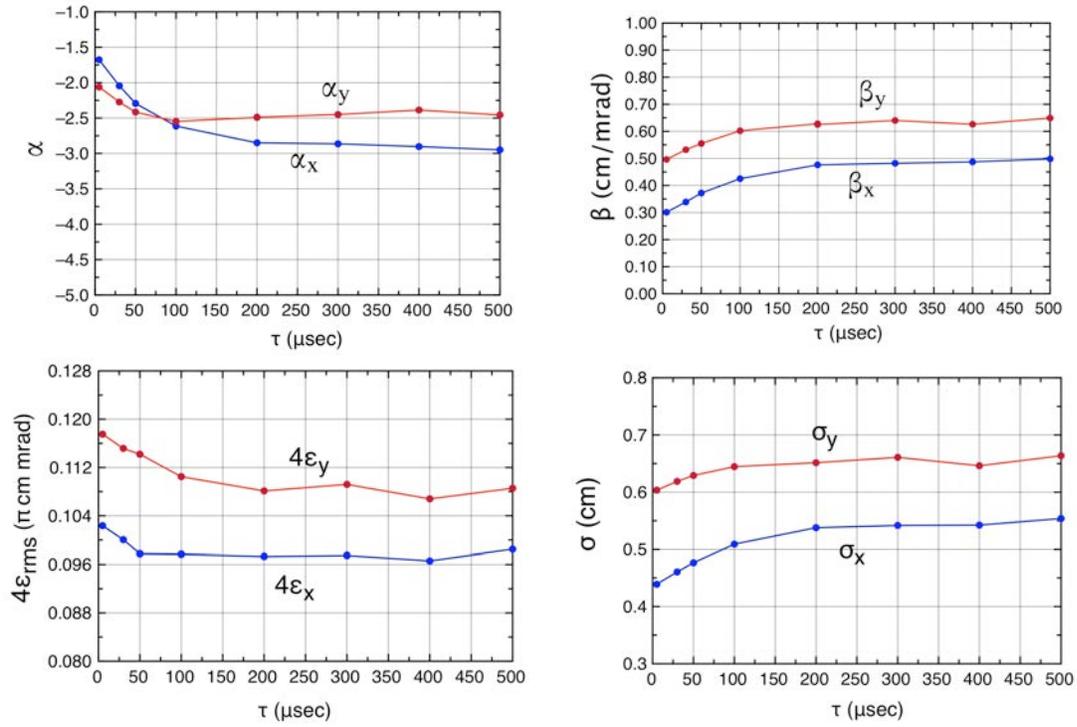

Figure 6: Parameters of 750-keV H- beam at TBEM4 as functions of beam pulse.

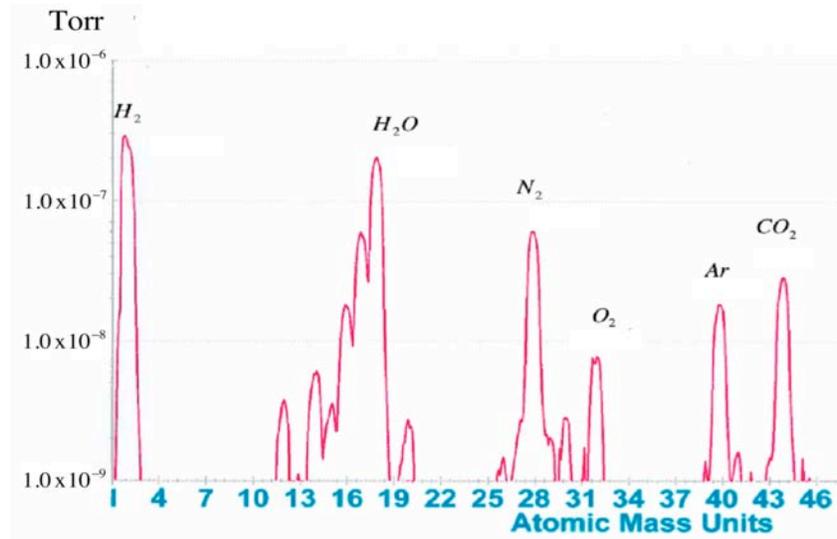

Figure 7: Residual gas analyzer scan.

Figure 8. Generation of macroparticle distribution from TBEM1 emittance scan.

Figure 9. TBEM2 beam emittance scan and results of BEAMPATH run between TBEM1 and TBEM2 emittance stations with effective current $I_{eff} = 0.56\ I_{beam}$.

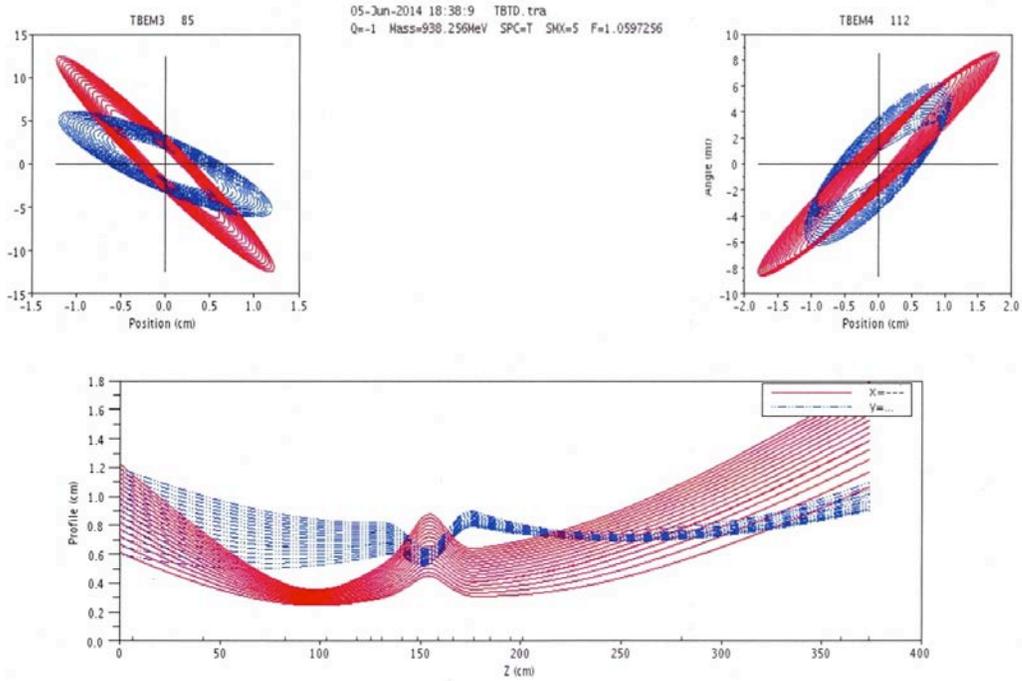

Figure.10. TRACE simulation between TBEM3 – TBEM4 with constant current and variable emittance.

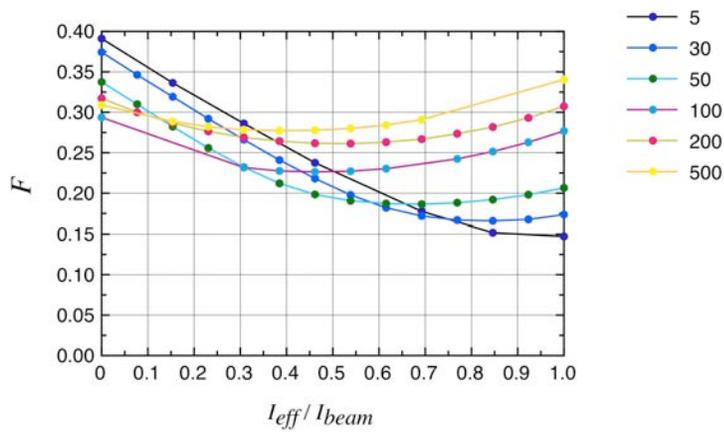

Figure 11: Mismatch factor $F$ as a function of effective beam current in BEAMPATH simulations between TBEM4 and TDEM1 (numbers indicate pulse length in µs).

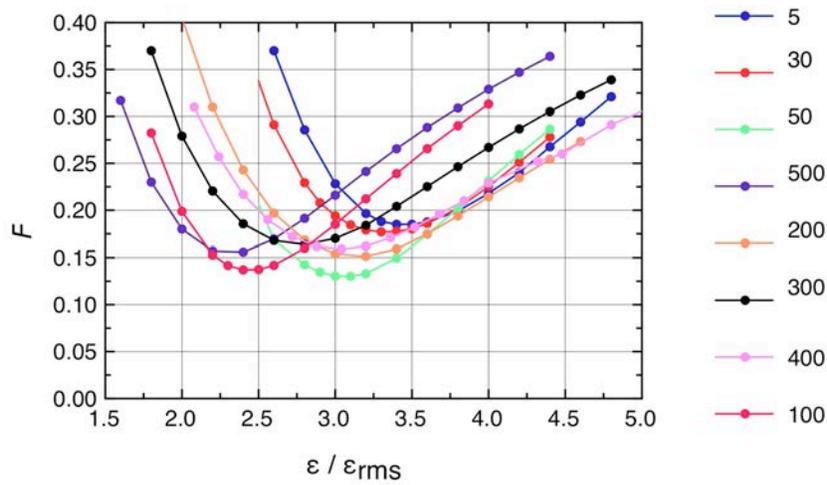

Figure 12: Mismatch factor $F$ as a function of effective beam emittance in TRACE simulations between TBEM4-TDEM1 (numbers indicate pulse length in μs).

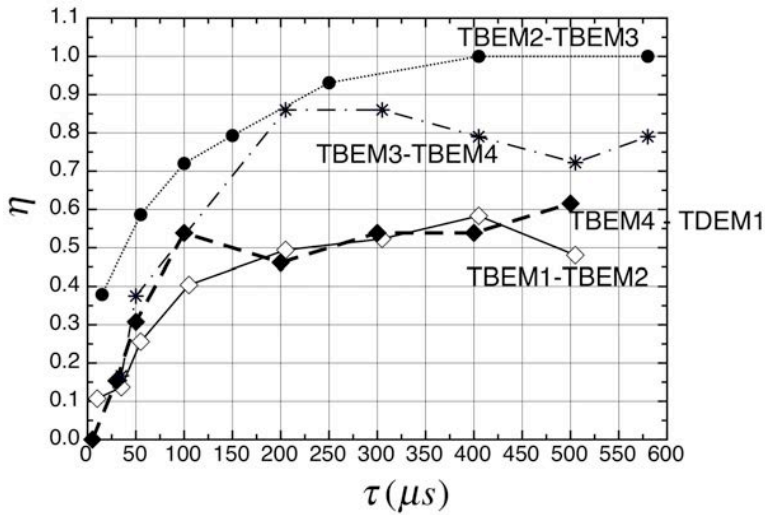

Figure 13: Space charge neutralization as a function of pulse length along the channel.

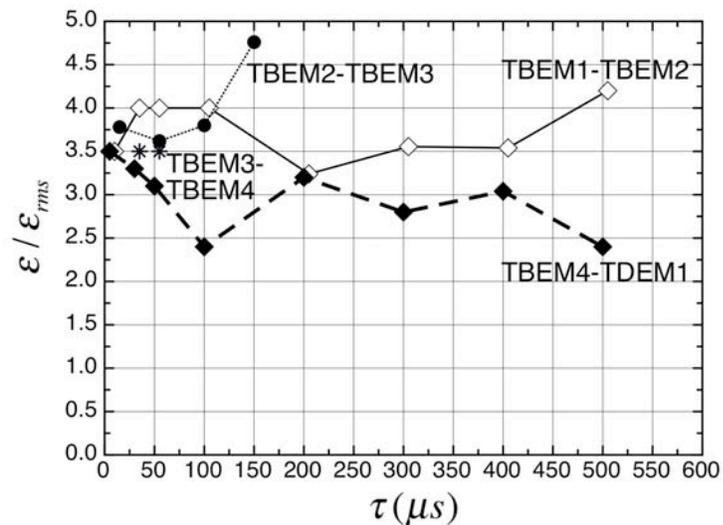

Figure 14: Effective beam emittance as a function of pulse length along the channel.

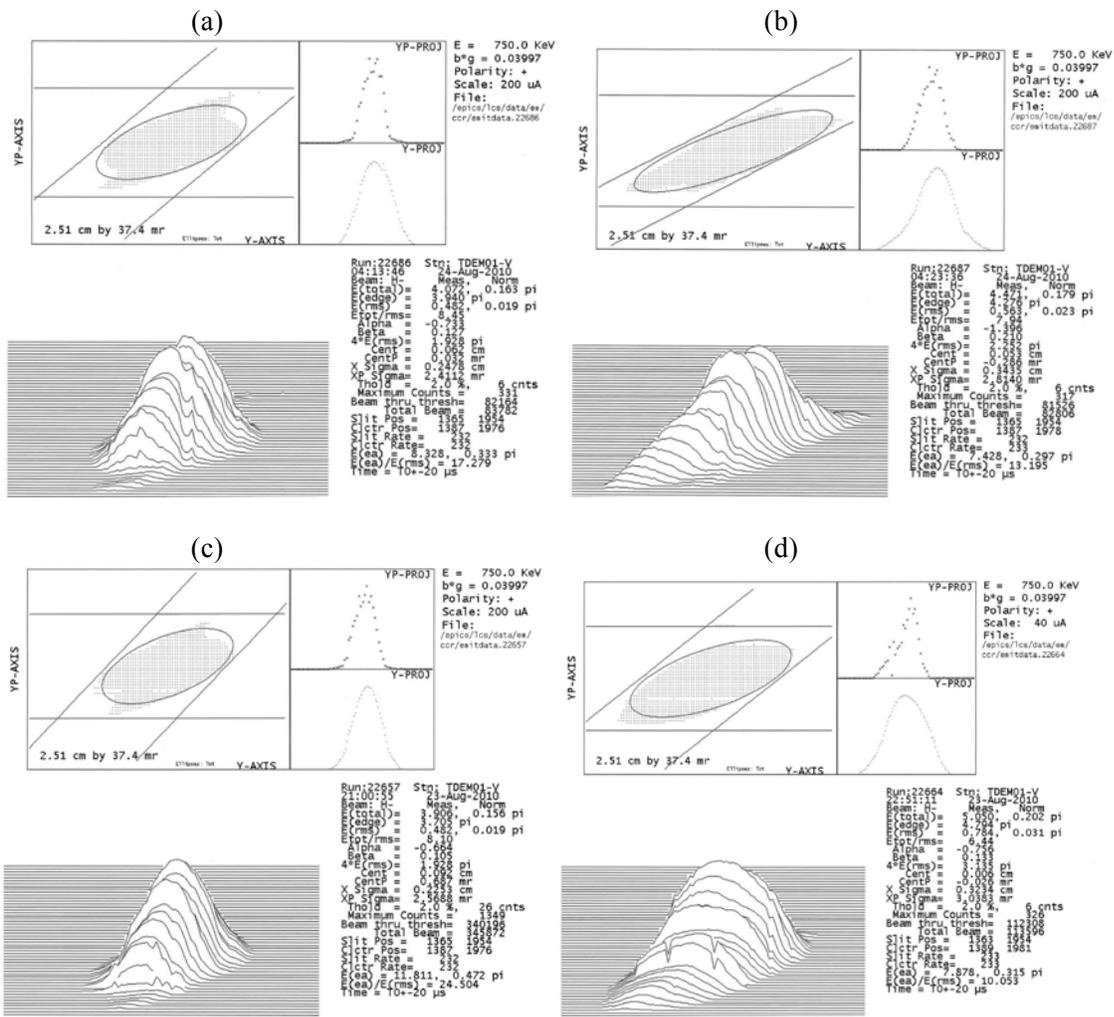

Figure 15: Effect of RF bunching and chopper pulse length on beam emittance at TDEM1:
(a) bunchers off, chopper off, (b) bunchers on, chopper off,
(c) buncher off, chopper pulse 290 ns, (d) buncher off, chopper pulse 36 ns.

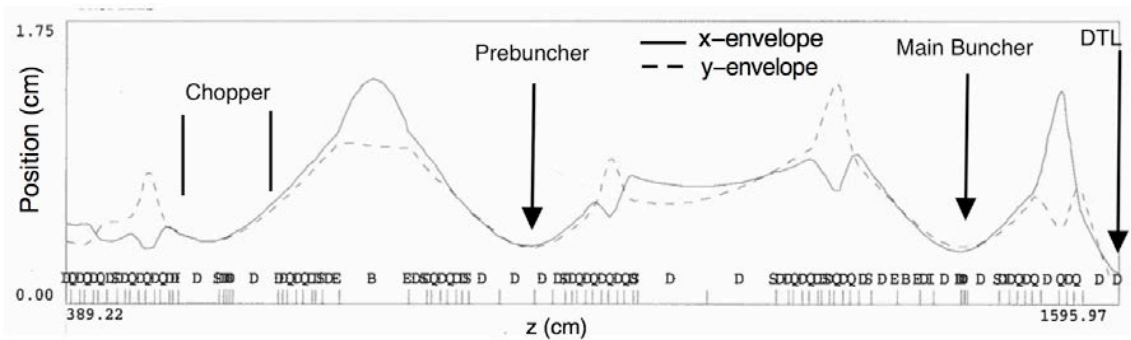

Figure 16: TRACE calculations of matched beam envelopes along the beamline: (solid) horizontal, (dotted) vertical.